\documentclass[journal]{IEEEtran}
\usepackage{blkarray}
\usepackage[cmex10]{amsmath}
\usepackage{amssymb}
\usepackage{arydshln}
\usepackage{amsmath,bm}
\usepackage{graphicx}
\usepackage[numbers,sort&compress]{natbib}
\usepackage[overload]{empheq}
\usepackage{color,soul}
\usepackage{float}

\usepackage{nomencl,etoolbox}
\makenomenclature
\renewcommand\nomgroup[1]{%
  \item[\bfseries
  \ifstrequal{#1}{C}{Constants}{%
  \ifstrequal{#1}{I}{Indices}{%
  \ifstrequal{#1}{M}{Matrices}{%
  \ifstrequal{#1}{S}{Sets}{%
  \ifstrequal{#1}{V}{Variables}{%
  \ifstrequal{#1}{Z}{Other Symbols}{}}}}}}%
]}

\renewcommand{\arraystretch}{1.5}

\newcommand{\abs}[1]{\left|#1\right|}

\begin{document}

\title{Fully Distributed DC Optimal Power Flow Based on Distributed Economic Dispatch and Distributed State Estimation}
%
%

\author{Qiao Li,~
        David Wenzhong Gao,~\IEEEmembership{Senior Member,~IEEE,}
        Lin Cheng,~
        Fang Zhang,~
        Weihang Yan
        }
\maketitle

\begin{abstract}
Optimal power flow (OPF) is an important technique for power systems to achieve optimal operation while satisfying multiple constraints. The traditional OPF are mostly centralized methods which are executed in the centralized control center. This paper introduces a totally Distributed DC Optimal Power Flow (DDCOPF) method for future power systems which have more and more distributed generators. The proposed method is based on the Distributed Economic Dispatch (DED) method and the Distributed State Estimation (DSE) method. In this proposed scheme, the DED method is used to achieve the optimal power dispatch with the lowest cost, and the DSE method provides power flow information of the power system to the proposed DDCOPF algorithm. In the proposed method, the Auto-Regressive (AR) model is used to predict the load variation so that the proposed algorithm can prevent overflow. In addition, a method called constraint algorithm is developed to correct the results of DED with the proposed correction algorithm and penalty term so that the constraints for the power system will not be violated. Different from existing research, the proposed method is completely distributed without need for any centralized facility.
\end{abstract}


%
\IEEEpeerreviewmaketitle

\section{Introduction}
\IEEEPARstart{O}{ptimal} Power Flow is a technique used in power system control center to achieve the optimal operation under certain constraints. Similar to the economic dispatch problem, most OPF methods minimize the operating cost of power systems. Additionally, some power flow related constraints are also considered. In conventional OPF methods, the optimization is usually performed by the centralized control center. However, in recent years, with the development of renewable energy and smart grid, power systems become more and more distributed. For these highly distributed power systems, it is not efficient to collect all data to one centralized place for computation. So, the distributed OPF method is needed for these distributed power systems. Also, with the distributed OPF method, the communication network of the power system can be designed as distributed, which is more robust than centralized communication network.

In this paper, the distributed method is developed with the DCOPF (DC Optimal Power Flow), which is an OPF method based on the DC power flow analysis. Compared with ACOPF (AC Optimal Power Flow), DCOPF provides a good-enough result for most power system applications with much higher speed. This paper addresses the DDCOPF problem considering the line flow constraints. There are some existing papers for distributed OPF problem. In \cite{2014Erseghe}, a distributed OPF algorithm is developed, in which the alternating direction multiplier method (ADMM) is used to decompose the optimization problem into several subproblems. However, in this method, the power system is required to be divided into several areas, and the centralized method is still used within each area. So the algorithm is not completely distributed. On the other hand, only the local power constraint and voltage limit constraint are considered in \cite{2014Erseghe}, but the constraints related to multiple buses, such as line flow constraint, are not discussed. It makes the proposed method in \cite{2014Erseghe} less useful in practice, since many important problems in OPF are multi-bus problems, e.g., the line flow constraint and the N-1 contingency problem. Similarly, three DDCOPF methods based on ADMM are proposed in \cite{2016Wang}, in which the power system is also divided into small areas, and the centralized method is also needed within each areas. The distributed OPF is also discussed in \cite{hug2009decentralized,nogales2003decomposition,dall2013distributed,bakirtzis2003decentralized,biskas2005decentralized} . These papers are also based on area partitioning method which is not fully distributed.
In \cite{kargarian2015distributed}, the authors proposed a Distributed Security-Constrained Unit Commitment (DSCUC) which is closely related to distributed OPF. The paper uses analytical target cascading
(ATC) method to achieve the DSCUC. However, this method also requires the system to be partitioned into several sub-areas.

The previous studies show that the totally distributed OPF is hard to be realized. The reason is that the power flow analysis is hard to be decomposed since the calculation of power flow requires the information across the entire power system (or at least a part of the system). In order to achieve the completely distributed OPF, a new OPF method is developed in this paper based on the distributed economic dispatch (DED) method \cite{DistEDPaper} and the distributed state estimation (DSE) method \cite{DistSEPaper}. In this scheme, the proposed method includes two algorithms. One is running in the smart meters and another is executed by the controllers of the generators. The smart meter algorithm is based on the DSE algorithm which provides the power flow information for the generators. The algorithm for the generators has two parts, i.e., the DED algorithm and the constraint algorithm. The DED method is employed to optimize the generation dispatch in the power system. The constraint algorithm ensures the optimized results from the DED method to satisfy the line flow constraints. This paper will mainly focus on the design of the constraint algorithm and how to integreate it with the DED method and DSE method to realize the DCOPF. The constraint algorithm corrects the optimized results with two approaches. In the first approach, called correction algorithm, the constraint algorithm first predicts the operating point of the power system and then corrects the movement of the operating point to restrict it within the feasible region. In the second approach, called penalty term, if the constraints are already violated, the algorithm will use a penalty term to pull the operating point back to the feasible region.
Since the DED and DSE methods are totally distributed approaches, and the constraint algorithm has no requirement for the centralized facilities, then the proposed OPF method is completely distributed.

The major contributions of this paper are listed as follows:
\begin{itemize}
\item This paper proposes a totally distributed OPF method which has never been realized before. In comparison, distributed OPF methods reported in existing literature are not truly distributed, since the centralized method is still utilized inside each region or area of the power system.

\item Compared to the local constraints in the previous papers \cite{2014Erseghe,DistEDPaper}, the multi-bus constraints, i.e., the line flow constraint, in the OPF problem is considered in our paper. So, the proposed method is much more useful in practice than the previous methods.

\item The basic concept of the proposed constraint algorithm is to restrict the operating point of the power system within the feasible region so that the line flow constraints can be satisfied. This idea can be adopted in future distributed OPF method with the consideration of other constraints, e.g., contingency constraints \cite{powerGOCBook}.
\end{itemize}


This paper is organized as follows:
In section \ref{Section: Preliminary}, the preliminary knowledge about the DED method and the DSE method are presented. In section \ref{Section: Proposed Approach}, the DDCOPF method is proposed. At the first part of this section, the framework of the proposed DDCOPF method is presented. In order to prevent the overflow in the transmission lines, the new method to predict and check the overflow in power system is developed based on AR model in subsection \ref{Section: PredictAndCheck}. Then, the correction algorithm and penalty term are introduced in the following subsections. In section \ref{Section: Simulation Results}, two simulation cases to verify the proposed DDCOPF method in a 39-bus test system are provided. Finally, section \ref{Section: Conclusion} concludes this paper and introduces future work.

\section{Preliminary}\label{Section: Preliminary}

\subsection{Distributed Economic Dispatch \cite{DistEDPaper}}
Economic dispatch (ED) is a method to optimize the generation assignment of generators to reach the lowest operational cost of the power system. The DED is a method to realize the economic dispatch in a distributed manner. Suppose that there are totally $n_b$ buses in the power system, $m_l$ transmission lines linking those buses, and $n_g$ generator buses (Assuming that each generator bus has one generator. But the conclusion will not change if there are multiple generators or loads on the bus). The ED problem can be described as follows \cite{powerGOCBook},
\begin{subequations}\label{FunED}
\begin{align}
\text{Min}~ F(P)=&\sum_{i=1}^{n_g} F_i(P_{g,i})\label{objFunED}\\
s.t. ~~~~~~\sum_{i=1}^{n_g} P_{g,i}&=P_{load}+P_{loss}\label{balConstED}\\
P_{g,i}^{min}\le &P_{g,i}\le P_{g,i}^{max}\label{outputConstED}
\end{align}
\end{subequations}
where $P_{g,i}$ is the power reference of generator $i$. ${F_i}(P_{g,i})$ is the cost function of the $i$th generator. $P_{load}$ is the total load in the power system. $P_{loss}$ is the power system's loss. $P_{g,i}^{min}$ and $P_{g,i}^{max}$ are the lower and upper active power output limits of the generator $i$, respectively. Typically, the generation cost function in (\ref{objFunED}) is a quadratic function as following,
\begin{equation}\label{costFun}
  F_i(P_{g,i})=\alpha_i+\beta_i P_{g,i}+\gamma_i P_{g,i}^2
\end{equation}
where $\alpha_i$, $\beta_i$ and $\gamma_i$ are the coefficients of the cost function for the $i$th generator.
\nomenclature[V]{$P_{g,i}$}{power reference of generator $i$}
\nomenclature[V]{$P_{load}$}{total real power load of the power system}
\nomenclature[V]{$P_{loss}$}{total loss of the power system}
\nomenclature[C]{$P_{g,i}^{min}$}{the minimum output of the $i$th generator}
\nomenclature[C]{$P_{g,i}^{max}$}{the maximum output of the $i$th generator}
\nomenclature[C]{$\alpha_i$,$\beta_i$,$\gamma_i$}{parameters of the cost function of $i$th generator}
\nomenclature[V]{$n_b$}{number of buses}
\nomenclature[V]{$n_g$}{number of generator buses}
\nomenclature[V]{$m_l$}{number of transmission lines}

According to the ED method, assuming that $P_{g,i}$ is within the limits, the optimal solution of the objective function can be obtained by setting the references of the generators such that the incremental cost of all generators are equal,
\begin{equation}\label{nessCondED}
  \frac{dF_i(P_{g,i})}{dP_{g,i}}=\lambda~~~~~,\text{for}~i=1,2,...,n_g
\end{equation}
where $\lambda$ is the Lagrange multiplier.

\nomenclature[V]{$\lambda$}{Lagrange multiplier}

Based on paper \cite{DistEDPaper}, in a distributed system, the Lagrange multiplier (or incremental cost) can be calculated by the following DED algorithm,
\begin{subequations}\label{distED}
\begin{align}[left = {\empheqlbrace}]
  &\lambda_i (k)=\lambda_i (k\!-\!1) + d\lambda_i(k)\label{lamConsen}\\
  &d\lambda_i(k)=\tau\!\sum_{j\in \mathcal{N}_i} w_{ij}\left(\lambda_j(k\!\!-\!\!1)-\lambda_i(k\!\!-\!\!1)\right) + \Delta\lambda_i(k)\label{lamConsen2}\\
  &\Delta\lambda_i(k) = 2\gamma_i\left[K\!p_i \Delta^2\!f_i(k) +\tau K\!i_i \Delta{f_i(k)}\right] \label{goalAttrc}\\
  &\Delta^2 f_i(k)=\Delta f_i(k)-\Delta f_i(k\!-\!1)\label{ddFreq}\\
  &\Delta f_i(k) = f_0 - f_i(k) \label{dFreq}
\end{align}
\end{subequations}
where $\lambda_i (k)$ is the estimated Lagrange multiplier on $i$th generator at time $k$. $\tau$ is the time interval between two iterations. $w_{ij}$ is a weight between buses $i$ and $j$ determining the convergence speed. $Kp_i$ is the proportional gain and $Ki_i$ is the integral gain. $f_i(k)$ is the frequency measured by the $i$th generator at time $k$. $f_0$ is the rated frequency (60Hz or 50Hz).

\nomenclature[V]{$\lambda_i (k)$}{estimated $\lambda$ by the $i$th generator at time $k$}
\nomenclature[C]{$K\!p_i$,$K\!i_i$}{respectively, proportional gain and integral gain of the $i$th generator}
\nomenclature[V]{$f_i(k)$}{the frequency measured by the $i$th generator at time $k$}
\nomenclature[C]{$f_0$}{the rated frequency (60Hz or 50Hz)}

As shown in (\ref{lamConsen}) and (\ref{lamConsen2}), the consensus protocol is employed to drive the estimated Lagrange multipliers of all generators to a certain value, which results in an optimal condition.
In addition, equations (\ref{goalAttrc}, \ref{ddFreq}, \ref{dFreq}) behave like a PI frequency controller to ensure that the power balance constraint (\ref{balConstED}) is satisfied.
Finally, for the output constraint (\ref{outputConstED}), the generation reference can be limited when the reference is calculated with the Lagrange multiplier as follows,
\begin{equation}\label{solutionED}
P_{g,i}^*(k)=
    \begin{cases}
        \displaystyle{\frac{\lambda_i (k)-\beta_i}{2\gamma_i}} &, \text{if}~P_{g,i}^{min}< P_{g,i}^* (k)< P_{g,i}^{max}\\
        P_{g,i}^{min} &,\text{if}~P_{g,i}^* (k) \le P_{g,i}^{min}\\
        P_{g,i}^{max} &,\text{if}~P_{g,i}^* (k) \ge P_{g,i}^{max}
    \end{cases}
\end{equation}
where $P_{g,i}^*(k)$ is the solution of the ED problem at time $k$.
\nomenclature[V]{$P_{g,i}^*(k)$}{desired power reference of generator $i$ at time $k$}

\subsection{Distributed State Estimation \cite{DistSEPaper}}
In traditional power system, state estimation is a widely implemented technique \cite{powerGOCBook,powerGOCBook2013,sun2013analog} which is usually performed by a centralized facility, e.g. SCADA (Supervisory Control And Data Acquisition). The distributed state estimation is a method to estimate the states of power systems without centralized facility. In order to realize the distributed state estimation, an information propagation algorithm is proposed in paper \cite{DistSEPaper} as follows,
\begin{equation}\label{infoProp}
  \dot{x}_i(k)=I^0_i \cdot \sum_{j\in \mathcal{N}_i} \left(x_j(k)-x_i(k)\right)
\end{equation}
where $I^0_i$ is a $n\times n$ diagonal matrix whose diagonal elements are all 1 except a 0 at the $i$th entry, i.e. $I^0_i\triangleq\textbf{Diag}([1, 1, ..., 1, 0, 1, ..., 1])$; $x_i(k)$ is the information, e.g., the measurement data, from the $i$th node.

According to the information propagation algorithm (\ref{infoProp}), the distributed state estimation algorithm is developed in \cite{DistSEPaper} as shown in the equations (\ref{distSE}),
\begin{subequations}\label{distSE}
\begin{align}[left = {\empheqlbrace}]
&Z'_{i,j}(k)=Z_{i,j}(k) \text{, for}~ j\neq i\\
&Z'_{i,j}(k)=z_{i,j}(k) \text{, for}~ j=i \label{inputZiPower}\\
&Z_i(k\!+\!1)=Z'_i(k)+\tau I^0_i \sum_{j\in \mathcal{N}_i}w_{ij}^Z\left(Z'_j(k)-Z'_i(k)\right)\label{infoEstOutPower}\\
&\hat{\theta}_i(k+1)=\left(H^T R^{-1} H\right)^{-1}H^\text{T}R^{-1}Z_i(k+1)\label{stateEstFinalPower}
\end{align}
\end{subequations}
where $Z_i(k)$ denotes the estimated measurement values of all the nodes in the system by the node $i$ at time $k$. $\hat{\theta}_i(k)$ is the estimated state values on the node $i$ at time $k$. $R$ and $H$ are the covariance matrix and the observation matrix, respectively.

The first equation (\ref{inputZiPower}) in the algorithm describes the input of the algorithm, where $Z'_{i}(k)$ is a modification of $Z_{i}(k)$ whose $i$th entry is replaced by the local measurement $z_{i,i}(k)$ at time $k$. The equation (\ref{infoEstOutPower}) is the information propagation algorithm with weight coefficient $w_{ij}^Z$. In addition, the equation (\ref{stateEstFinalPower}) represents the distributed estimation with least square method \cite{DistSEPaper,GraphBook}. The initial values of the algorithm are: $Z_{i,i}(0)=z_{i,i}(0)$; $Z_{i,j}(0)$ can be arbitrary number (e.g. zero), where $Z_{i,j}(k)$ denotes the $j$th entry of the vector $Z_i(k)$.

\section{Proposed Approach}\label{Section: Proposed Approach}
In the proposed distributed OPF method, the power system has two networks: the power network and the communication network. The power network connects the loads and generators together with the transmission lines. The communication network links all the generator controllers together, and also connects all the smart meters together. In addition, each generator controller connects to a smart meter (or, a smart meter can be installed as a part of the controller) to obtain the power flow information. In the proposed scheme, the DSE algorithm runs on the smart meters, and the proposed DDCOPF algorithm is executed on the generator controllers.

\subsection{The idea and framework of the DDCOPF}
In the OPF problem, the optimization considers not only the economic dispatch problem with the power balance constraint (\ref{balConstED}) and the generation limit constraint (\ref{outputConstED}), but also some power flow related constraints. For example, this paper focuses on the line flow limit constraints (\ref{lineLimitConstScalar}) as follows \cite{powerGOCBook2013},
\begin{equation}\label{lineLimitConstScalar}
  \frac{1}{x_{ij}} \abs{\theta_i-\theta_j} \leq{P_{ij}^{max}} ,~~~~\text{for}~i,j=1,2,...,n_b;~i\neq j
\end{equation}
where $x_{ij}$ is the reactance between bus $i$ and bus $j$. $\theta_i$ and $\theta_j$ are the voltage phase angles on the bus $i$ and $j$, respectively. $P_{ij}^{max}$ is the power flow limit for the transmission line between bus $i$ and bus $j$. According to the DC power flow \cite{powerADBook}, the constraint (\ref{lineLimitConstScalar}) for the entire power system can also be written in the vector form,
\begin{equation}\label{lineLimitConstVector}
\abs{P_f}=\abs{H\theta}=\abs{H(B')^{-1}P}\leq{P_f^{max}}
\end{equation}
where $P_f$ is the vector of the line flow in the power system. $P_f^{max}$ is the vector of line flow limits. $H$ is the matrix to convert the phase angle values to the power flow values. $\theta$ and $P$ are the vectors of phase angles and nodal injections for buses $2, 3, ...,n_b$, respectively.  $B'$ is called ``B-prime" matrix, which transfers the phase angles into the power injection \cite{powerGOCBook}. Since bus $1$ is the reference bus and its phase angle is always zero, the $B'$ matrix does not include the row and column for bus $1$, so the impact of power injection at bus $1$ cannot be directly calculated with the $B'$ matrix. Therefore a new matrix $T$ is used in this paper to calculate the power flow. The $T$ is constructed as
\begin{equation}\label{matrixT}
T=\left[\!
    \begin{array}{cccc}
      -B_{12} & -B_{13} & ... & -B_{1,n_b}\\
       B_{22} & -B_{23} & ... & -B_{2,n_b}\\
      -B_{32} &  B_{33} & ... & -B_{3,n_b}\\
       \vdots & \vdots  & \ddots & \vdots\\
      -B_{n_b,2} & -B_{n_b,2} & ... & B_{n_b,n_b}
    \end{array}\!
\right]^g
\end{equation}
\nomenclature[M]{$T$}{linear transformation from power injections to phase angles}
where, $[\!\!~*\!\!~]^g$ denotes the generalized inverse of the matrix, so $T$ is a $(n_b\!\!-\!\!1)\!\times\! n_b$ matrix, and
\begin{equation}
B_{ij}=
    \begin{cases}
        \displaystyle{0} &,\text{if buses $i$ and $j$ are not connected}\\
        \displaystyle{\frac{1}{x_{ij}}} &,\text{if buses $i$ and $j$ are connected}\\
        \displaystyle{\sum_{\substack{k=2\\ k\neq i}}^{n_b\!-\!1}\! {B_{ik}}} &,\text{if}~i=j
    \end{cases}
\end{equation}
Now, the inequality constraint (\ref{lineLimitConstVector}) becomes,
\begin{equation}\label{lineLimitConstNew}
\abs{P_f}=\abs{H\theta}=\abs{HTP}\leq{P_f^{max}}
\end{equation}

\begin{figure*}
  \centering
  \includegraphics[width=5.5in]{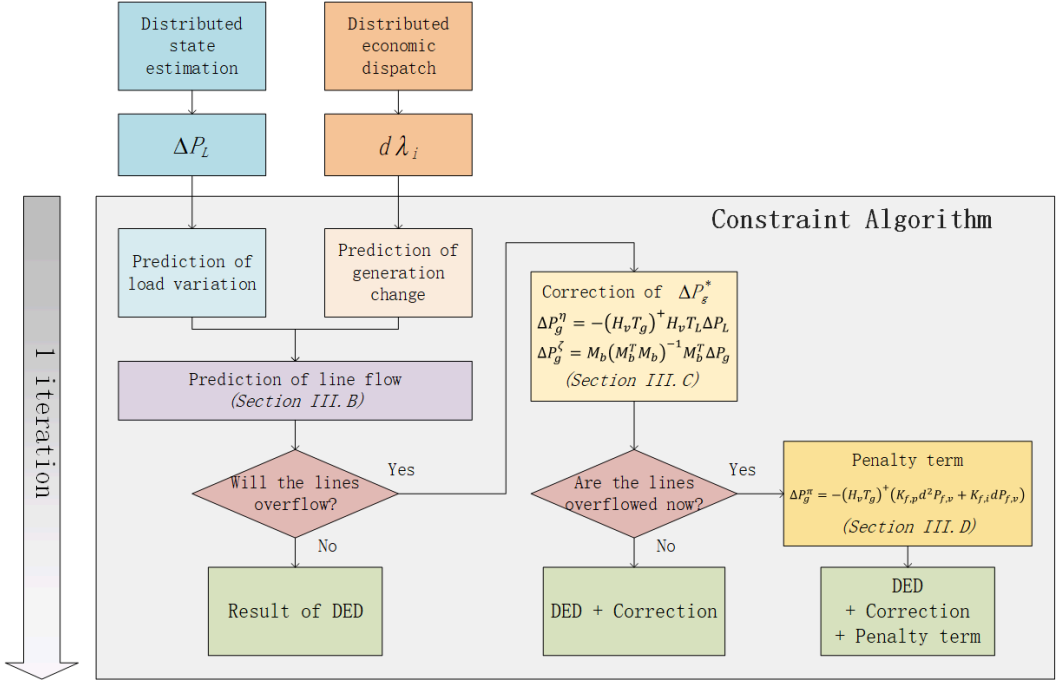}
  \caption{The flow chart of the proposed constraint algorithm}\label{mainFlowChart}
\end{figure*}

To solve the DDCOPF problem, the DED method \cite{DistEDPaper} is employed in this paper to constantly update the solution of the ED problem (\ref{FunED}) to approach the optimal dispatch. However, the constraint (\ref{lineLimitConstNew}) defines a feasible region for the solution of the DDCOPF problem. In order to keep the constraint (\ref{lineLimitConstNew}) being satisfied, the DDCOPF should limit the solution $P_{g}$ of the DED algorithm within the feasible region. In this paper, a line flow constraint algorithm (as shown in Fig.\ref{mainFlowChart}) is proposed to correct the solution of the DED method, so that the final solution of the DDCOPF is restricted in the feasible region.

According to Fig.\ref{mainFlowChart}, at the first, the information from the DED algorithm and DSE algorithm are adopted to predict the line flows of the next step in the power system and determines whether there will be overflow. If no potential overflow in the power system, the generation references calculated by the DED algorithm is directly applied on the generators. However, If overflow will happen on any line, the correction algorithm will be performed to correct the references of the generators to prevent the overflow. In addition, if the overflow has already happened in the system, a penalty term will be added to the references to reduce the line flow. In the following subsections, the different parts in the framework in Fig.\ref{mainFlowChart} will be described.

\subsection{Line Flow Prediction and Overflow Checking}\label{Section: PredictAndCheck}

This section discusses the method to predict and check the overflow in the power system. The method is based on the DC power flow model, and the  Auto-Regressive (AR) model is adopted for the load prediction.

In the DC power flow model, due to the linearity, the load buses and the generator buses in the power system can be considered separately as follows,
\begin{equation}\label{lineFlowCalGandL}
P_{f}=H\theta=HTP=H(T_g P_g+ T_L P_L)
\end{equation}
where $P_g$ and $P_L$ are the power injection vector of the generator buses and load buses, respectively. $T_g$ is a matrix constructed according to (\ref{matrixT}) but with only the generator buses. Similarly, $T_L$ is developed by considering only the load buses.

Before the time step $k$, the prediction for the line flow at time $k$ can be made by (\ref{lineFlowPredGandL}).
\begin{equation}\label{lineFlowPredGandL}
\begin{aligned}
P_{f}(k)\! =&H\left(\theta(k-1)+\Delta\theta(k)\right)\\
=&P_{f}(k-1)\!+\!H T_g\Delta\!P_g(k) \!+\! H T_L\Delta\!P_L(k)
\end{aligned}
\end{equation}
The power flow $P_f(k-1)$ at the previous step can be obtained from the DSE algorithm. So, the object is to predict the update of the power injection vector $\Delta P_g(k)$ and $\Delta P_L(k)$.

\nomenclature[V]{$P_{f}(k)$}{vector of line flow of all lines at time $k$}

The update of the generation reference vector $\Delta P_g(k)$ can be obtained from the result of DED algorithm. In (\ref{lamConsen2}), the update of the Lagrange multiplier $d\lambda_i(k)$ is calculated. According to (\ref{solutionED}), the final result of the generation reference is computed based on the Lagrange multiplier $\lambda_i(k)$. So, for the controller of the $j$th generator, the update of the generation reference $\Delta P^j_{g,i}(k)$ of the $i$th generator can be estimated by (\ref{genRefUpdatePred}) with its local Lagrange multiplier $\lambda_j(k)$, since $\lambda_j(k)=\lambda_i(k)$ after the consensus protocol is converged.
\begin{equation}\label{genRefUpdatePred}
\begin{aligned}
\Delta\!P^j_{g,i}&(k)=P_{g,i}^*(k)-P_{g,i}^*(k\!-\!1)\\
=&\begin{cases}
\displaystyle{\frac{d\lambda_j(k)}{2\gamma_i}}\!\!\!&, \text{if}~P_{g,i}^{min} \!<\!\! P_{g,i}^* (k) \!<\!\! P_{g,i}^{max}\\
0&, \text{otherwise}
\end{cases}
\end{aligned}
\end{equation}

Therefore, by storing the parameters $\beta_i$ and $\gamma_i$ for $i=1,2,...n_g$ in each generator controller, the prediction of the generation reference update vector $\Delta\!P_g(k)=[\Delta P_{g,1}^j(k); \Delta P_{g,2}^j(k); ...; \Delta P_{g,n_g}^j(k)]$ can be obtained by the equation (\ref{genRefUpdatePred}). Note that the notation $\Delta\!P_g(k)$ here is a vector for the $j$th generator. But, since the equation is the same for all controllers and for the sake of simplicity, the superscript $j$ is omitted.

Now, let's consider the prediction of the load variation $\Delta P_L(k)$. There are many methods to predict the load in power system \cite{2011Fan,2012Singh}. In this paper, an Auto-Regressive (AR) model of order 2 is used to predict the load variation as follows,
\begin{equation}\label{ARmodel}
\Delta P_L(k)=\varphi_1 \Delta P_L(k\!-\!1) + \varphi_2 \Delta P_L(k\!-\!2)
\end{equation}
the parameters $\varphi_1$ and $\varphi_2$ can be calculated based on historical data of $\Delta P_L(t)$ by Least Square method in (\ref{ARmodelParaLS}),
\begin{equation}\label{ARmodelParaLS}
\text{Min}{\sum^T_{t=3}\left({\Delta P_L(t)-\varphi_1 \Delta P_L(t\!-\!1) - \varphi_2 \Delta P_L(t\!-\!2)}\right)^2}
\end{equation}

By the AR model, the prediction $\Delta P_L(k)$ can be made by (\ref{ARmodel}) with the previous updates $\Delta P_L(k-1)$ and $\Delta P_L(k-2)$, whose previous updates are provided by the DSE algorithm.

After the prediction of generation reference update $\Delta P_g(k)$ and prediction of load variation $\Delta P_L(k)$ are obtained, the prediction for the line flow $P_{f}(k)$ can be calculated by (\ref{lineFlowPredGandL}). Then, the overflow in the power system can be checked, and the results are put in a column vector $O\!F(k)$ whose each entry is defined by,
\begin{equation}\label{overFlowCheck}
O\!F_u(k)=
\begin{cases}
1 ,&P_{f,u}(k)\geq P^{max}_{f,u}\\
0 ,&\text{otherwise}
\end{cases}
\end{equation}
where $O\!F_u(k)$ is the $u$th entry of the vector $O\!F(k)$ at time $k$; $u=1,2,...,m_l$. $P_{f,u}(k)$ is the line flow in the line $u$ at time $k$. $P^{max}_{f,u}$ is the maximal line flow allowed for the line $u$.

\subsection{Correction of Power References Update $\Delta P_g^*$}
According to Fig.\ref{mainFlowChart}, the power references of the generators calculated by the DED algorithm are directly output if the prediction shows no overflow in the power system in the next step. However, if any overflow will happen, the power references should be adjusted to avoid the violation of the line flow constraint. In this section, a method to correct the power references is proposed to make the solution of the DDCOPF stay within the feasible region.

\begin{figure}
  \centering
  \includegraphics[width=2in]{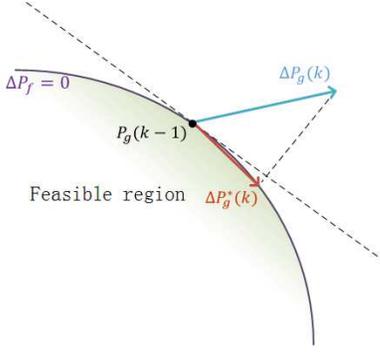}\\
  \caption{The correction of the power reference update vector}\label{correctRef}
\end{figure}

Since the power balance constraint (\ref{balConstED}) and the output limit constraint (\ref{outputConstED}) are already satisfied by the DED algorithm, the correction algorithm in this section should adjust the result of the DED algorithm to meet the line flow constraint (\ref{lineLimitConstNew}).
Assuming that the current solution $P_{g}(k-1)$ of the DED algorithm is inside the feasible region but near the boundary as shown in Fig. \ref{correctRef}. If the line flow prediction (\ref{lineFlowPredGandL}) shows that the constraint (\ref{lineLimitConstNew}) will be violated at time $k$, i.e. the solution $P_{g}(k)$ will be out of the feasible region, then it is necessary to adjust the update $\Delta P_g(k)$ of the DED solution to prevent the further increasing of the power flow on the lines, denoted by $P_{f,v}(k)$, which will be overflowed according to the prediction. This adjustment can be realized by letting the increment $\Delta P_{f,v}(k)$ to be zero as follows,
\begin{equation}\label{noChangeOfPf}
\Delta P_{f,v}(k) \!=\! H_v(k) \left(T_g \Delta\!P_g(k) \!+\! T_L \Delta\!P_L(k)\right)=0
\end{equation}
where $H_v(k)$ is the matrix related to the lines which will violate the constraint, and it is defined as follows,
\begin{equation}\label{calHv}
H_v(k)= \textbf{Diag}([OF_1(k),OF_2(k),...,OF_{m_l}(k)])\cdot H
\end{equation}

\nomenclature[V]{$H_v(k)$}{the matrix related to the lines which will violate the constraint at time $k$}

In (\ref{noChangeOfPf}), since $\Delta P_g(k)$ is the adjusted update for the DED solution which should be calculated, the equation (\ref{noChangeOfPf}) can be rewritten into the nonhomogeneous form as follows,
\begin{equation}\label{nonHomoEq}
H_v(k)T_g \Delta\!P_g(k) = - H_v(k) T_L \Delta\!P_L(k)
\end{equation}


Because $H_v(k)T_g$ is usually not a full ranked square matrix, the solution cannot be simply computed by its inverse. So, according to linear algebra \cite{algebraArtin}, the solutions of this nonhomogeneous equation are the sum of a particular solution $\Delta P^\eta_g(k)$ of the corresponding nonhomogeneous equation and the general solutions $\Delta P^\zeta_g(k)$ of the homogeneous equation $H_v(k)T_g \Delta\!P_g(k)=0$.
Here, by equation (\ref{nonHomoEq}), it is convenient to consider that the particular solution $\Delta P^\eta_g(k)$ is to counteract the effect of the load variation $\Delta\!P_L(k)$ on the line flow. Then, the solution $\Delta P^\zeta_g(k)$ of the homogeneous equation is the desired power reference update for the power system with constant loads.

First, let us solve the particular solution $\Delta P^\eta_g(k)$ of the nonhomogeneous equation. Since the particular solution is to neutralize the effect of the loads, the difference (if it is not zero) between the left-hand side (the line flow caused by generators) and the right-hand side (the line flow caused by loads) of the equation (\ref{nonHomoEq}) is required to be minimized. In other words, the solution should be the least square solution. Also, it is better to minimize the norm of the solution $\Delta P^\eta_g(k)$, so that the result is more energy efficient. Therefore, the Moore-Penrose pseudoinverse \cite{linearAlgebra} is used to solve the particular solution $\Delta P^\eta_g(k)$ as in (\ref{calNonHomoParSol}), since it gives the least square solution with minimal norm.
\begin{equation}\label{calNonHomoParSol}
\Delta\!P^\eta_g(k) = -(H_v(k)T_g)^+ H_v(k)T_L \Delta\!P_L(k)
\end{equation}
where $\Delta P_L(k)$ comes from the prediction (\ref{ARmodel}). The superscript notation $(*)^+$ denotes the Moore-Penrose pseudoinverse.


Next, let us consider the general solution of the homogeneous equation $H_v(k)T_g \Delta\!P_g(k)=0$. It is true that all the solutions of the homogeneous equation form the kernel of the matrix $H_v(k)T_g$. Since the kernel is a subspace of the $\mathbb{R}^{n_g}$, it is possible to find a basis that spans the kernel. Suppose that a basis matrix $M_b$ is defined in which its column vectors are the basis of the kernel, and the coordinates for the vectors in the kernel are denoted by $X$. Because any vector in the kernel satisfies the homogeneous equation, then $X$ can be any value. Note that the matrices $M_b$ and $X$ are related to $H_v(k)$, hence they are not constant value, i.e. $M_b(k)$ and $X(k)$. Therefore, the general solution of the homogeneous equation is,
\begin{equation}\label{HomoGenSol}
\Delta\!P^\zeta_g(k) = M_b(k) X(k)
\end{equation}
So, the homogeneous equation becomes,
\begin{equation}\label{HomoGenNew}
H_v(k)T_g M_b(k) X(k)=0
\end{equation}

\nomenclature[V]{$M_b$}{basis matrix of the solution space of the homogeneous equation}

As discussed above, the homogeneous equation describes the power system with constant loads, thus the change of line flow is only related to the generation. Since the object is to correct the original power reference updates to meet the requirement, it is necessary to preserve as much information from the original solution $\Delta\!P_g(k)$ as possible, and also satisfy the homogeneous equation (\ref{HomoGenNew}). So, it is natural to use Least Square method to find the solution $X$ in which the corrected vector $\Delta\!P^\zeta_g(k)$ is closest to the original vector $\Delta\!P_g(k)$ and is in the kernel. According to the the equation (\ref{HomoGenSol}), the Least Square solution is,
\begin{equation}\label{HomoGenLSSol}
\Delta\!P^\zeta_g(k)=M_b(k)(M_b^\text{T}(k) M_b(k))^{-1} M_b^\text{T}(k) \Delta\!P_g(k)
\end{equation}
Now, as the particular solution of the nonhomogeneous equation and the Least Square solution of the homogenous equation are obtained, the desired solution of the nonhomogeneous equation (\ref{nonHomoEq}) is
\begin{equation}\label{nonHomoGenSol}
\begin{aligned}
\Delta\!&P^*_g(k) = \Delta\!P^\eta_g(k) + \Delta\!P^\zeta_g(k)\\
=& -(H_v T_g)^+ H_v T_L \Delta\!P_L  + M_b (M_b^\text{T} M_b )^{-1} M_b^\text{T} \Delta\!P_g
\end{aligned}
\end{equation}
here, to make the equation clear, the time variable $k$ is omitted. $\Delta\!P^*_g(k)$ is the corrected power reference update with inputs $\Delta\!P_g(k)$ and $\Delta\!P_L(k)$.

\subsection{Penalty Term for the infeasible operating point}
Ideally, the correction algorithm above can ensure the solutions of the DDCOPF method to stay in the feasible region. However, in practice, some factors may cause the actual operating point (the power flow) of the power system to be infeasible.

In the correction equation (\ref{calNonHomoParSol}), the load variation $\Delta P_L(k)$ is estimated by the AR model (\ref{ARmodel}), thus the error of the predication will be propagated into the final result. Consequently, the solution $\Delta P_g(k)$ may not be restricted on the boundary and drift out of the boundary due to the prediction error. Also, since the change of power flow is eliminated by the correction algorithm by letting $\Delta P_f(k)=0$, the infeasible result cannot return to the feasible region.

In addition, the inertia of the generator causes the delay between the output reference (the solution of the DDCOPF) and the actual output. So, when the reference is bounded on the boundary of the feasible region, the actual output may not stop increasing instantly, but overshoot beyond the limit. Then, due to the equation $\Delta P_f(k)=0$, the operating point may not go back to the feasible region.

Due to the above reasons, if the operating point moves out of the feasible region, a penalty function is needed in the constraint algorithm to pull the infeasible operating point back to the feasible region. In other words, the power flow of the overflowed lines should be decreased to satisfy the line flow constraint (\ref{lineLimitConstNew}). According to the equation (\ref{noChangeOfPf}), the power flow increment on the overflowed lines $\Delta P_{f,v}$ includes two parts such that $\Delta P_{f,v}=H_v T_g \Delta P_g+H_v T_L \Delta P_L$. Due to the correction algorithm, the loads related term $H_v T_L \Delta P_L$ is eliminated by the solution $\Delta\!P^\eta_g(k)$ of the nonhomogeneous equation (\ref{nonHomoEq}). Also, instead of fixing the operating point by letting its increment $\Delta P_{f,v}=0$ in the correction algorithm, the penalty term should decrease the power flow by making the increment $\Delta P_{f,v}$ to be negative as follows,

\begin{equation}\label{decreasePowerFlowEq}
\Delta P_{f,v} = H_v(k)T_g \Delta\!P^\pi_g(k) = -F_s(k)
\end{equation}

where $F_s(k)$ is a vector with positive entries which represents the step size of the power flow decreasing. $\Delta\!P^\pi_g(k)$ is the change added on the generation reference to produce the penalty.
Due to the inertia of the power system, the change $\Delta\!P^\pi_g(k)$ on the generation reference cannot affect the power flow instantly, so it is impossible to reduce the power flow $P_{f,v}$ to the limit $P_{f,v}^{max}$ in one iteration by directly letting $\Delta P_{f,v}=-(P_{f,v}-P_{f,v}^{max})$.
Therefore, the concept of PI controller is employed to calculate the step vector $F_s$ to gradually decrease the power flow to the feasible region.

\begin{equation}\label{PIdecreasingStep}
F_s(k)=K_{f,p} d^2P_{f,v}(k)+K_{f,i} dP_{f,v}(k)
\end{equation}

where $dP_{f,v}(k)=P_{f,v}(k)-P_{f,v}^{max}$ and $d^2P_{f,v}(k)=dP_{f,v}(k)-dP_{f,v}(k-1)$. $K_{f,p}$ and $K_{f,i}$ are the proportional gain and integral gain, respectively. Therefore, according to the equations (\ref{decreasePowerFlowEq}) and (\ref{PIdecreasingStep}), the change of generation reference to produce the penalty is obtained with the pseudoinverse as follows,

\begin{equation}\label{changGenRefPenalty}
\Delta\!P^\pi_g \!=\! -(H_v T_g)^+(K_{f\!,p} d^2P_{f\!,v} \!+\! K_{f\!,i} dP_{f\!,v})
\end{equation}

In the power system, when the operating point is out of the feasible region, this penalty term $\Delta\!P^\pi_g(k)$ can be added to the result from the correction algorithm, so that the final result becomes $\Delta\!P^*_g(k) = \Delta\!P^\eta_g(k) + \Delta\!P^\zeta_g(k)+\Delta\!P^\pi_g(k)$.

\section{Simulation Results}\label{Section: Simulation Results}
\begin{figure}
  \centering
  \includegraphics[width=\hsize]{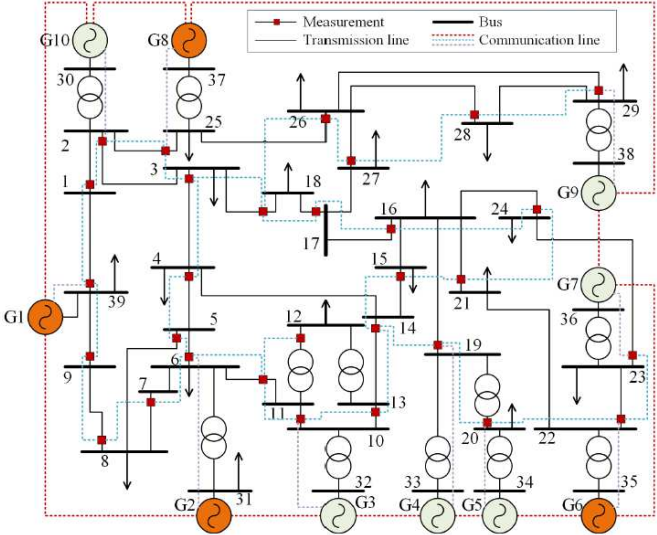}\\
  \caption{IEEE 39-Bus Power System}\label{39bus}
\end{figure}

A IEEE 39-Bus power system model with 10 generators and 39 buses is built in Matlab/Simulink as shown in Fig.\ref{39bus} \cite{pai2012energy}. The 39-Bus system is operated in 60Hz with 19 loads on different buses. The communication network in the power system includes two parts, one network connects all generator controllers together and another network links all smart meters in the system. In addition, each generator connects to a nearby meter to get the power flow information. As shown in Fig.\ref{39bus}, there is no centralized control center in the power system, so the system is totally distributed. The parameters of the generators are listed in Table \ref{genParameters39}.

\begin{table}[!t]
\renewcommand{\arraystretch}{1.3}
\caption{The Parameters of Generators in 39-bus system}
\label{genParameters39}
\centering
    \begin{tabular}{c||cccc}
      \hline
      ~ &{$\alpha_i$} &{$\beta_i$} &{$\gamma_i$} &$P_{rated}$~(MW) \\
      \hline
      DG1 & 561 & 8.08 & 0.00118 & 1000 \\
      DG2 & 310 & 7.8  & 0.00346 & 1000 \\
      DG3 & 278 & 7.85 & 0.00322 & 1000 \\
      DG4 & 453 & 8    & 0.00184 & 1000 \\
      DG5 & 453 & 8.1  & 0.00248 & 1000 \\
      DG6 & 524 & 7.95 & 0.00385 & 1000 \\
      DG7 & 384 & 7.86 & 0.00268 & 1000 \\
      DG8 & 368 & 7.75 & 0.00362 & 1000 \\
      DG9 & 572 & 8.12 & 0.00262 & 1000 \\
      DG10& 426 & 8.03 & 0.00368 & 1000 \\
      \hline
    \end{tabular}
\end{table}

\begin{figure}
  \centering
  \includegraphics[width=\hsize]{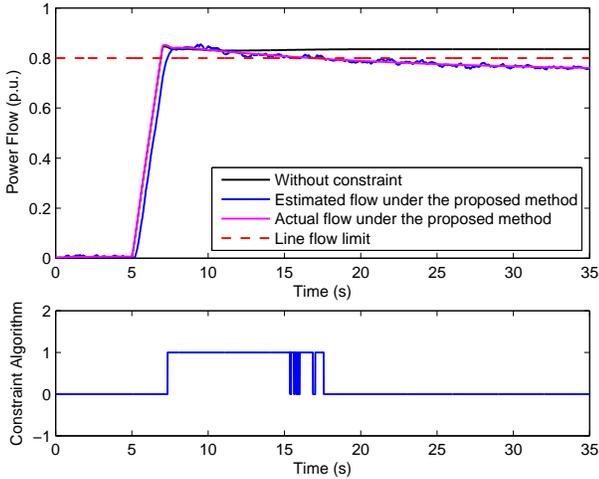}\\
  \caption{Case 1: The line flow on line 24 with 0.8 p.u. limit.}\label{case1line24}
\end{figure}

\begin{figure}
  \centering
  \includegraphics[width=\hsize]{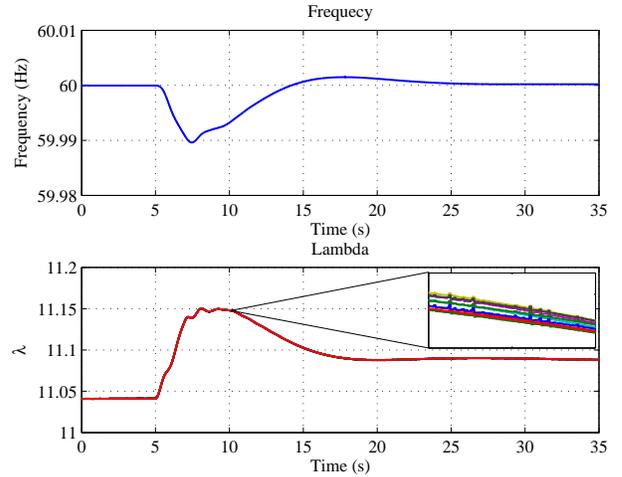}\\
  \caption{Case 1: The Lagrange multiplier and frequency of the system}\label{case1lambdaFreq}
\end{figure}

In this paper, two case studies are provided to show the performance of the proposed method. In the first case, the limit of the line flow on line 24 is set to 0.8 p.u. To simulate the load variation and overflow, the load on bus 24 increases by 100MW from 5s to 7s, so that the line 24 will overflow if no action is taken. The results are shown in Fig.\ref{case1line24} and Fig.\ref{case1lambdaFreq}.

The top figure in Fig.\ref{case1line24} shows that, without applying the proposed DDCOPF method, the line flow on line 24 will go beyond the limit and stays at about 0.82. But, with the proposed method, the line flow decreases down below the limit. In the figure, the estimated flow is from the smart meter with the DSE method. The estimation is almost the same as the actual power flow on the line, which means that the DSE method is pretty accurate. The sub-figure at the bottom of the Fig.\ref{case1line24} shows when the constraint algorithm is actived. In Fig.\ref{case1lambdaFreq}, the frequency of the system returns to 60Hz after the load variation and the Lagrange multipliers (Lambdas) are identical for all generators, which means that the optimal solution of the ED problem with power balance constraint as described in equation (\ref{FunED}) is achieved.

\begin{figure}
  \centering
  \includegraphics[width=\hsize]{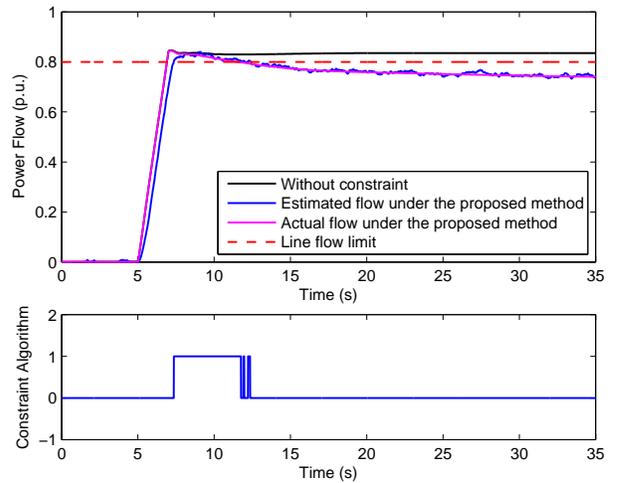}\\
  \caption{Case 2: The line flow on line 24 with 0.8 p.u. limit.}\label{case2line24}
\end{figure}

\begin{figure}
  \centering
  \includegraphics[width=\hsize]{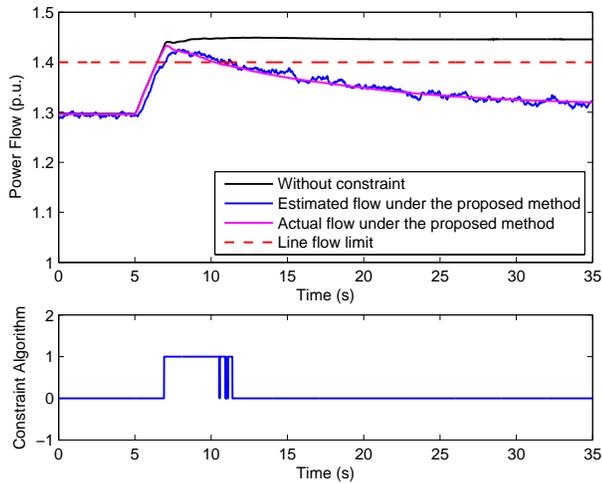}\\
  \caption{Case 2: The line flow on line 27 with 1.4 p.u. limit.}\label{case2line27}
\end{figure}

\begin{figure}
  \centering
  \includegraphics[width=\hsize]{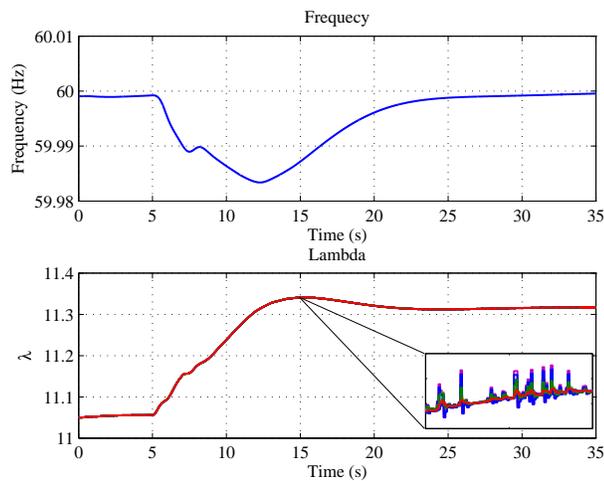}\\
  \caption{Case 2: The Lagrange multiplier and frequency of the system}\label{case2lambdaFreq}
\end{figure}

In another case study, the scenario in which two lines in the power system are overflowed at the same time is simulated. In this case, line 24 and line 27 exceed their maximal line flow, i.e., 0.8 p.u. for line 24 and 1.4 p.u. for line 27, after the load increasing on bus 24. The simulation results are shown in Fig.\ref{case2line24}, Fig.\ref{case2line27} and Fig.\ref{case2lambdaFreq}. Similar to the previous case, the line flow on both line 24 and line 27 decrease to the feasible region. Also, the frequency and Lagrange multiplier show that the system operates in the optimal condition.

The simulation is run on a PC with Intel i7-7700HQ 2.8GHz CPU and 8GB memory. The time consumed by each iteration of the algorithm is less than 0.000002s. So, the algorithm does not require a very powerful computation device and is valid for real-time running.

\section{Conclusion}\label{Section: Conclusion}
A DDCOPF method is proposed in this paper to address the OPF problem with line flow constraints. The proposed method is completely distributed without the need of the centralized control center as in the traditional method. The proposed method is developed based on the DED method and the DSE method. The DED is to find the optimal solution of the OPF. The auto-regressive model is employed to identify the potential overflow in the power system. Then, the constraint algorithm consisting of correction algorithm and penalty term is proposed to force the solution of OPF to satisfy the line flow constraints. The proposed method is simulated in a 39-Bus system model in Matlab/Simulink. The results show that the constraint algorithm can limit the line flow within the feasible region while reaching the optimal operation.



%

\ifCLASSOPTIONcaptionsoff
  \newpage
\fi



%
\bibliographystyle{IEEEtran}
\bibliography{ref}
\end{document}